# A framework for understanding data science
Michael L. Brodie mlbrodie@seas.harvard.edu
Data Systems Laboratory, School of Engineering and Applied Sciences
Harvard University, Cambridge, MA USA
DRAFT February 16, 2024

**Abstract**

The objective of this research is to provide a framework with which the data science community can understand, define, and develop data science as a field of inquiry. The framework is based on the *classical reference framework[1] (axiology, ontology, epistemology, methodology)* used for 200 years to define knowledge discovery paradigms and disciplines in the humanities, sciences, algorithms [28], and now data science. I augmented it for automated problem-solving with *(methods, technology, community) [3][4].* The resulting *data science reference framework* is used to define the *data science knowledge discovery paradigm* in terms of the *philosophy of data science* addressed in *[3]* and the *data science problem-solving paradigm, i.e., the data science method[2]*, and the *data science problem-solving workflow*, both addressed in this paper. The framework is a much called for [8][33] *unifying framework* for data science as it contains the components required to define data science. For insights to better understand data science, this paper uses the framework to define the emerging, often enigmatic, *data science problem-solving paradigm* and *workflow,* and to compare them with their well-understood scientific counterparts – *scientific problem-solving paradigm* and *workflow.*

## 1. Need to better understand inscrutable data science.

Our 21st C world is transforming faster than ever before in human history. Understanding and utilizing knowledge in our knowledge-based, digital world is increasingly challenging due to its expanding scope (e.g., bioinformatics, neurology, astrophysics), scale (e.g., medical knowledge doubles every 73 days [7]), and complexity (neurology studies the most complex phenomenon known to man). Technology not only contributes to this growth with automation, but it also provides data science knowledge representation and discovery capabilities that are transforming every digitally expressible human endeavor for which there is adequate data. Data science is used to understand (reason over) existing knowledge and to discover new knowledge (i.e., solve problems, generate innovations) at previously impossible scopes, scales, complexity, and power, often beyond human capacity to understand. We do not yet know how or what AI-based methods[3] *learn* in training nor *infer* in analysis yet offer the potential of solving otherwise insoluble problems including existential problems like climate change and cancer, with equal potential of causing harm.

We offer a framework with which to better understand data science to achieve at least two things. First, that it enables more efficient, effective, and beneficial knowledge discovery and innovation with properties such as accuracy, integrity, and robustness. Second, that it aligns with positive human values and life in two risk scenarios: *the bad actor problem* of AI used to cause harm; and *the alignment problem* to ensure that the goals of an AI align with positive human values.

---

[1] The classical reference framework has evolved over 200 years and is not attributable to any one individual.

[2] We use one of many useful definitions of science and of data science, as explained in Appendix §8.1.

[3] Appendix §8.2 defines conventional and AI-based data science.

AI-based data science is currently based on AI-based methods such as Machine Learning (ML), Deep Learning (DL), neural networks (NN), dynamic programming (DP), Generative AI, and Large Language Models (LLMs). While essential data science reference framework components, e.g., epistemology of AI-based methods, are inscrutable, the framework aids in developing a comprehensive a definition of data science to contribute to identifying, defining, and addressing open research challenges.

To better understand AI-based data science, this paper uses the framework to gain valuable insights into data science by comparing data science with science by elaborating their respective *problem-solving paradigms* and corresponding *workflows*. Two types of insight emerge. The first arises from comparing the emerging, inscrutable *data science knowledge discovery paradigm* with the well-known and well-defined *scientific knowledge discovery paradigm.* The second derives from identifying fundamental differences between data science and science, our previously most powerful and best understood knowledge discovery paradigm. For example, science provides certainty with provable, incomplete knowledge of the natural world, at human scale; in contrast, data science provides uncertain, probabilistic knowledge of any phenomenon for which there is adequate data, often at scales beyond human understanding. *The uncertainty of data science may better reflect reality than the perceived certainty of science*.

While *what is data science*? may seem philosophical and far from urgent, practical concerns, it must be understood to maximize potential benefits, to identify and minimize risks, and to anticipate our 21st C world. Such *data science thinking* [13] is required to gain insights from data science problem-solving. Our objectives are motivated by applications and research challenges exemplified below.

### 1.1. *Motivating data science applications*

Comparing data science with science strongly suggests that due to its scope, scale, complexity, and power, the already widely deployed *AI-based data science will replace science as our dominant knowledge discovery paradigm*. Consider example applications with positive and negative potential that motivate the need to better understand AI-based data science.

### 1.1.1. Economics motivates widespread adoption

Matt Welsh, CEO of Fixie.ai [38] estimated the cost of a human software engineering day in Silicon Valley in 2023 to be $1,200. Welsh estimated the cost of an equivalent amount of work produced by an AI agent, e.g., Fixie or Copilot, to be $0.12. The economics will lead all programming achievable with AI to done by AI, replacing millions of human programmers. However, like all such AI applications, this is just the beginning. While AI programming works well for modules, it has yet to be applied to large scale programming, e.g., query optimizers, DBMSs, and enterprise applications, but that too seems to be inevitable. Consider another benefit of AI programming. A human programmer may take weeks to discover all APIs relevant to a programming task, to determine which is best, and how to deploy it. An AI agent can learn every API and its uses then select and deploy the best one in seconds. Methods are being developed to keep such an AI API agent current automatically and cheaply compared with humans having to repeat the entire task for each API choice.

### 1.1.2. AI will dramatically improve medical care

Centuries of research and practice have produced a vast amount of medical knowledge that is the basis of medical education. Doctors are trained with this knowledge then by practical experience. First, "It is estimated that the doubling time of medical knowledge in 1950 was 50 years; in 1980, 7 years; and in 2010, 3.5 years. In 2020 it is projected to be 0.2 years — just 73 days" [7]. Doctors cannot stay current, but an AI medical agent can, automatically. Second, doctors practical experience is limited to the patients that they treat while an AI agent can learn from every available patient medical history. These advantages have been realized in AI-based medical care agents, e.g., to improve maternal health, and are being applied in many medical care and mental health applications. Due to the critical nature of such knowledge, AI agents are being deployed as doctor's assistants that offer the most relevant knowledge, assessments, and treatment plans that leverage all relevant and recent knowledge.

[Delfina care](#) is an example of data-driven digital care systems closing existing gaps in pregnancy care by enabling earlier intervention and monitoring. Isabel Fulcher, Delfina Chief Scientific Officer, describes in a [Delfina Care lecture](#) how the Delfina team leverages data to design data visualizations to improve remote patient monitoring, deploy predictive algorithms to enable personalized interventions, and implement quasi-experimental designs to optimize patient care experiences and outcomes using causal reasoning.

### 1.1.3. The world will accept AI as it did photography

Photography emerged in the early 19th C transforming the creation of images from skilled, knowledge- and innovation-based, costly human endeavors that produced high-value images, to an automated activity requiring little skill, knowledge, or time, significantly impacting the value of images. Over time, that fundamental change was accepted and integrated into modern life. In the 21st C, knowledge and innovation can be automated based on existing data, knowledge, and data science, making knowledge innovation and discovery fast, cheap, and indistinguishable from human counterparts. Previously, the resulting knowledge was practically and legally considered an asset or property, i.e., IP. How will the automation of knowledge and innovation change our world?

### 1.1.4. AI vs. human products: indistinguishable, more flexible

Since its launch in London in 2022, sold-out performances of ABBA Voyage earns $2M per week. This performance of ABBA, the 1970s, four member Swedish rock group, has been seen by over 1M people. The performers are AI avatars of the septuagenarian musicians. Soon, the performances may be in cities around the world with the avatars speaking in the local languages with the correct lip and body movements. Another example is that AI can produce books so rapidly that "Amazon banned authors from self-publishing more than three e-books a day on its Kindle platform and required publishers to clearly

label books by robot writers." [The Guardian, Sept. 20, 2023] "Now and Then", a song written and recorded in the late 1970's by John Lennon, was re-produced using AI to improve the original audio and to mix in Paul McCartney and Ringo Star voices, despite Lennon having passed in 1980. These are examples of human endeavors that are being automated with AI, often indistinguishable from their human counterparts.

## 1.2. Motivating data science challenges

While the essential aspects of AI-based data science learning and discovery are inscrutable, framing data science and its research challenges may lead to better understand specific features of AI-based knowledge discovery. Consider three such examples.

### 1.2.1. Power versus safety

There appears to be a *fundamental tradeoff* between 1) maximizing the scope, scale, complexity, and power of unconstrained, untrained AI-based knowledge discovery methods, (trained) models, solutions, and results, and 2) limiting those properties by constraining models, problems, solutions, and results with guardrails to bound their behavior for safety and accuracy. This poses significant research challenges for achieving an optimal tradeoff between maximizing those properties and operating within required bounds. We need to understand the tradeoff, related challenges, and research to address them.

### 1.2.2. Data quality

Data quality is a measure of the extent to which training or inference data is an accurate, true characterization of a phenomenon, often for an intended purpose. More precisely, *data quality* is a measure of the extent to which a data value for a property of a phenomenon satisfies its specification. While verifying the correspondence between a value and its specification can be automated, the critical correspondence is between the value and the reality that it represents, verifiable only by humans. This must be done at scale.

### 1.2.3. Data scale

How much data is required to address the above two challenges – to maximize the scope, scale, and complexity of AI-based data science, and to achieve human-evaluated veracity of training and inference data, trained models, solutions, and results. Can there be too much or too little data? Can we measure whether a vast dataset has a distribution that accurately and adequately reflects reality? If not, how do we deal with its inherent biases? Adequate training data challenges involve, at least 1) data at scale, and 2) biased data.

- **Data at scale**: Scale is currently a challenge as the largest datasets, e.g., the Internet, have been exhausted. This will change as the Internet grows expanding beyond its predominantly white, western origins and as massive, yet unused private and public data stores become available, e.g., from private and public institutions, governments, businesses, and individuals. This will result in a data marketplace with complex ownership, legal, and governance issues yet to be addressed. Synthetic data is being used, but it may not represent reality and the extent to which it does represent reality, it may not reflect the natural occurrences, i.e., distributions.
- **Biased data** True, accurate data that reflects the values and knowledge of the data source e.g., culture, enterprise, individuals, reflects their inherent biases. Models trained on such data reflect those biases. For example, medical models trained on white, western patients may adequately reflect the medical issues of the source patients but may not adequately reflect those of non-white, non-western patients. Medical applications may require that data quality requirements include accurately reflecting the target patient population.

- **Eliminate societal biases to improve AI models** There is considerable research on improving the aspirational quality of data used for training AI-based models, e.g., eliminating systemic racial bias from AI-based models used in medicine – medical diagnosis and treatment that currently discriminate against people that are not white, English speaking, or live in the Western world. This means that real data cannot be used as is. Would the ideal training dataset for a specific medical diagnosis represent all people equitably relative to their existence in the population? Should real data be curated so to remove or correct data that is erroneous or incorrectly represents the intended phenomena?

In summary, core data science data challenges are 1) *data representation* – accurately representing features of phenomena, 2) *data purpose* – accurately and adequately representing data for specific analytical purposes, 3) *data adequacy and scale*, and 4) *verifying truth* – means with which the relevant community can evaluate data quality, hence, the quality of data science results [5].

### 1.3. Data science and science differ fundamentally

To better understand data science, we use the conventional problem-solving paradigm as a framework with which to compare the well-understood scientific problem-solving paradigm – *the scientific method* – with the emerging data science problem-solving paradigm – *the data science method*. As a context for the comparison, consider critical fundamental similarities and differences. Science and data science are knowledge discovery paradigms that employ the conventional problem-solving paradigm and workflow in fundamentally different ways reflecting their respective natures, introduced here, elaborated in §5.

One of the greatest innovations and resulting discoveries of the 21st C is AI-based data science that enables knowledge discovery, problem solutions, and innovation at *scopes (applications)*, *scales*, *complexity*, and *power* beyond those of science, often beyond human understanding [4]. Scientific analysis is restricted to the physical realm – measurable phenomena – and to problems at human scale – that humans can understand - at the cost of physical labor and time. In contrast, AI-based data science analysis operates in the digital realm automating and accelerating discovery for problems potentially at unfathomed scale with increasingly powerful, fast, and inexpensive computing and data resources.

AI-based data science introduces a fundamentally new knowledge discovery, generation, and problem-solving paradigm with methods that enable solutions not otherwise possible, not only at scopes, scales, and complexities never before possible, but for *any problem for which there is adequate data.* Central to data science is data at scale from which it learns (trains) and infers (discovers) patterns at scale and complexity. It knows nothing of human concepts, despite humans interpreting results as such. The results – discovered patterns – can be seen as phenomena that humans may have never imagined, hence potentially providing fundamentally novel conceptions of reality. Human concepts are artificial, i.e., conceived of by humans, as are data science-discovered innovations, created by AI-programmed computations. *Data-driven, AI-based data science will discover knowledge not previously discovered by science or conceived of by humans.*

Scientific experiments are governed by the scientific method that requires that experiments be validated, i.e., the experimental design is correct; the experiment was conducted according to the design; and the results verified, i.e., within acceptable tolerances of human hypothesized values. Validation and verification is done first by a scientific team and then authorized by the relevant community, e.g., journal, as required by the scientific reference framework. Hence, scientific problems, experiments, and results are well understood. In contrast, the nature, scope, scale, complexity, and power of AI-based data science enable previously unimagined benefits. However, these benefits render inscrutable AI-based data science (untrained) methods, (trained) models, and results (discovered knowledge, innovations) often posing major challenges. There are no theoretical or practical means for explanation – to explain and verify what or how methods learn or what or how models infer. Similarly, there are no theoretical or practical means for interpretation – to interpret and validate results in terms of the motivating domain problem. This poses two major risks. First, models and their results could be erroneous. A model may conduct the wrong analysis or conduct the right analysis incorrectly. This risk can be addressed to a limited degree within data science using empirically developed guardrails. Second, solutions and their results may be indistinguishable from their human produced counterparts. Whether correct or not, when applied in practice they may be destructive or cause harm, potentially at scale. These risks cannot be addressed within data science, prompting hyperbolic claims of risks of ending civil society or the world.

### 1.4. AI-based data science challenges

In ~500,000 papers [40], hundreds of thousands of researchers attempt to understand AI-based data science challenges such as 1) how to constrain AI with empirically developed guardrails, with considerable success; 2) how to train models to be more precise and less error prone, also with success; and 3) how AI solutions learn and infer, e.g., comparing AI-based learning versus human learning as a guide to alter AI-based algorithms, with limited success. These challenges fall into two areas – 1) understanding its reasoning (problem-solving paradigm), and 2) understanding the mappings between two sides of the conventional problem-solving paradigm (Fig. 1) – a) the motivating domain side – model, problem, solution, and result, and b) the analytical side – (AI-based) model, problem, solution, and result. The first challenge concerns the nature of AI-based reasoning. The second concerns the application of AI-based reasoning in problem-solving or knowledge discovery. This paper attempts to contribute to frame and understand these challenges.

### 3. Problem-solving paradigms and workflows

### 3.1. Conventional problem-solving paradigm

The conventional problem-solving paradigm (Fig. 1) is used to solve a problem concerning a phenomenon in a discipline, i.e., domain of discourse. Domain problems are normally solved directly on real world phenomena, by evaluating hypothesized solutions directly on those phenomena (left side of Fig. 1). Consider domain problem-solving on its own. A *domain problem* is expressed in the context of a known *domain model* as a hypothesized extension, parameterized by *domain problem parameters* over which the domain problem is evaluated. A *result specification* aids all problem-solving steps, especially the *domain problem result* that it specifies. A *domain solution* is created by augmenting the domain model with the hypothesized, parameterized domain problem. The domain solution is evaluated over the hypothesized *domain problem parameter values* against real *phenomena* producing a *domain solution result* that is then used to produce a *domain problem result*, i.e., the hypothesized solution is true or false.

Conventional problem-solving solves a domain problem aided by solving a corresponding, equivalent analytical problem that provides guidance for solving the domain problem and *vice versa*. Analytical problem-solving (right side of Fig. 1) uses an *analytical model* – typically a mathematical model of the phenomenon, possibly parameterized by *model parameters*, that has been demonstrated to correspond to the domain model via a domain-analysis map. A *domain-analysis map* aids expressing the domain problem as an *analytical problem* within an analytical model, that together with *analytical problem parameters* defines an *analytical solution*. A *result specification* is used to specify or bound the intended analytical problem result. The analytical solution can be applied by instantiating it with *analytical problem parameter values* and evaluating it over the *analytical data* that represents phenomenon instances thus solving the analytical problem and producing an *analytical solution result*. That result is used to evaluate whether the hypothesized analytical problem was true or false thus producing an *analytical problem result* that is verified against the result specification. The domain problem is solved only when the analytical problem result is interpreted in terms of the motivating domain problem producing the desired *domain problem result*. This is done using a domain-analysis map.

In summary, the *conventional problem-solving paradigm* consists of the domain problem-solving paradigm that is aided by the analytical problem-solving paradigm, and *vice versa*.

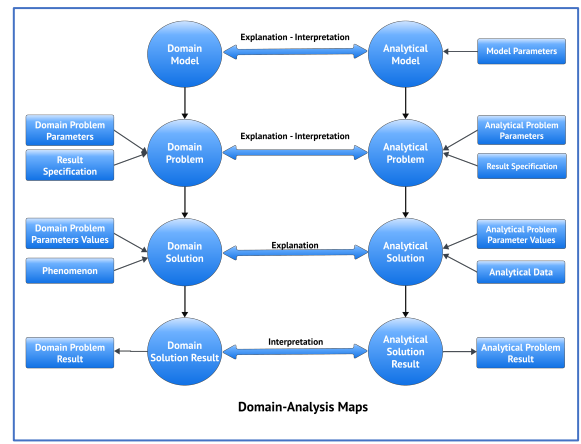

Figure 1: Conventional problem-solving paradigm.

In all problem-solving paradigms, domain-analysis maps establish correspondences between domain and analytical problem solving components at the same level. At a minimum, the correspondences aid in explaining, interpreting, and even deriving a domain component in terms of its mapped analytical component, or *vice versa*. Hence, they aid in developing and explaining problem-solving components at the same level, e.g., domain model to analytical model, as well as domain-analysis at different levels, e.g., a domain-analysis map between models aids developing a corresponding maps between problems. As described below, domain-analysis maps in science always establish equivalence; in data science they may provide insights.

*Derivation relationships* (downward arrows from upper to lower components in) are also valuable in all problem-solving paradigms. They represent theoretical and practical means for deriving and verifying lower from upper problem-solving components. Unlike a domain-analysis map used to establish a direct correspondence, each derivation relationship is specific to the problem and the discipline. For example, in the Higgs boson experiment (§3.2.2) the scientific model is the standard model or particle physics (SMPP) minus the Higgs boson. The scientific problem is the hypothesized behavior of the Higgs boson expressed as an SMPP extension.

### 3.2. Scientific problem-solving paradigm

This section describes the components of the *scientific problem-solving paradigm*[4] (Fig. 2), i.e., the conventional problem-solving paradigm applied in science. Scientific problem-solving is conducted on the domain side against real phenomena, i.e., by a scientific experiment, aided by analytical problem-solving on the analytical side, against data representing an equivalent problem in an independent, often mathematical or simulation, analytical model. The following sub-sections describe how the paradigm is applied in a scientific workflow.

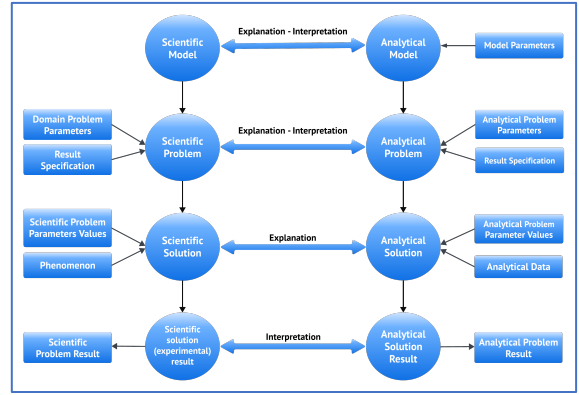

Figure 2: Scientific problem-solving paradigm.

Scientific problem-solving follows the scientific method to discover novel properties of a physical phenomenon for which there is a *scientific model*, expressed as a conceptual model, that represents the scientific knowledge of the phenomenon that has been proven theoretically and validated empirically. The purpose of scientific problem-solving is to solve a *scientific problem* defined as a hypothetical extension of the *scientific model* over *scientific problem parameters* against the real phenomena. A *result specification* characterizes a hypothesized scientific problem result including required precision and aids future problem-solving steps. The scientific model is augmented with the scientific problem and its scientific problem parameters to form, i.e., to derive and verify, a *scientific solution* (experiment). The experiment evaluates, i.e., measures, the phenomena[5] directly over the *scientific problem parameter values* producing a *scientific solution (experimental) result*, i.e., observations of the hypothesized values of properties of the phenomenon. The experimental result must be interpreted in terms of the motivating scientific problem to produce a *scientific problem result*. If the observed values satisfy the result specification, then the scientific solution, i.e., hypothesized behavior, is true; otherwise, it is false. Scientific truth is determined first by a successful experiment, supported by a corresponding analytical solution and ultimately by concurrence of the relevant scientific community. The community evaluates the model, problem, experiment, and results to authorize it by acceptance in a journal or a conference. Subsequently, the new scientific knowledge is curated, i.e., incorporated, into the scientific knowledge of the relevant discipline.

---

[4] There are many accepted definitions of the scientific problem-solving paradigm. The definition used here was chosen as a basis of comparison with the emerging data science problem-solving paradigm.

[5] Physical empiricism, or by analogy with data science terminology, learning from physical phenomena.

Analytical problem-solving is used to develop and guide scientific problem-solving. Consider the analytical problem solving components. Such an analytical solution provides means for designing, exploring (discovering), and evaluating analytical problems, solutions, and results.

An *analytical problem* parameterized by *analytical problem parameters,* hypothesizes the properties of the behavior of a phenomenon to be proven empirically on the scientific side. They are expressed as extensions of an *analytical model*, possibly parameterized by *model parameters*, that represents the scientific knowledge of the phenomenon that has been proven theoretically and validated empirically. An analytical model is an independent representation, often in mathematics or a simulation. Their equivalence is established with a verified domain-analysis map. The analytical side analyzes a model of the scientific side phenomenon.

An *analytical solution* is created, i.e., derived, by augmenting the analytical model with the hypothesized analytical problem parameterized over the conditions being explored by the *analytical problem parameters* that represents the hypothesized behavior of instances of the phenomenon. A *result specification* is used to specify the hypothesized analytical problem result and aids in all problem-solving steps. A verified domain-analysis map is used to map the analytical problems, solutions, and results to their scientific counterparts.

An analytical problem is solved (explored) by executing the analytical solution over specific *analytical problem parameter values* and *analytical data*, representing instances, and producing an *analytical solution result*, i.e., computational results of the hypothesized behavior of the phenomena. The analytical solution result must be interpreted in terms of the analytical problem to produce, i.e., discover, an *analytical problem result*. If that result satisfies the result specification, e.g., within the required bounds of the hypothesized properties of the behavior of the phenomenon, then the hypothesized analytical behavior is true; otherwise, it is false. Such analytical truth aids in establishing and documenting community authorization. Finally, the analytical problem result must be interpreted in terms of the motivating scientific problem result, and *vice versa*.

### 3.2.1. Scientific problem-solving workflow

Scientific problem-solving is conducted following the scientific method in the two phase scientific problem-solving workflow (Fig. 3) that uses the scientific and analytical scientific problem-solving paradigm components (Fig. 2) as follows and demonstrated in the next section with the Higgs boson experiment.

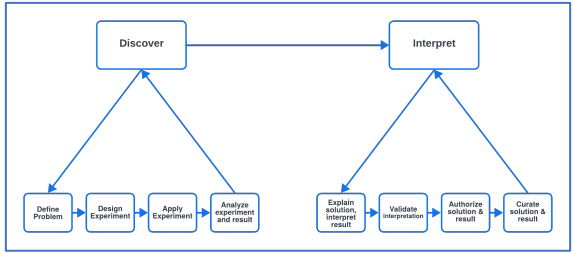

Figure 3: Scientific problem-solving workflow.

The *Discover* phase consists of four steps. In the *Define problem* step, a scientific problem with its scientific problem parameters is defined as a hypothetical extension of the relevant scientific model. A domain-analysis map is defined to express (interpret, explain) the scientific problem and model in terms of an analytical problem with its analytical problem parameters within its relevant analytical model with its model parameters. Result specifications are defined on the scientific and analytical sides to characterize the hypothesized scientific and analytic problem results and to aid in all problem-solving steps.

In the *Design experiment* step, an analytical solution is developed (derived) by augmenting the relevant analytical model with the hypothesized analytical problem parametrized by analytical problem parameters. The instances to which the analytical solution is to be applied are to be given as specific analytical problem parameter values and analytical data. At this point, the experiment designed on the analytical side is used to guide the configuration of the experiment on the scientific side. Previously defined domain-analysis maps are used to guide (derive) the definition of another domain-analysis map with which to map the analytical solution to the scientific solution (experiment) and its scientific problem parameter values and the phenomenon that is the subject of the experiment. Similarly, previously defined derivation relationships can be used to derive and verify the scientific solution. This step involves developing and configuring experimental apparatus with which to conduct the experiment designed on the analytical side, illustrated in §3.2.2 by the Large Hadron Collider (LHC) for the Higgs boson experiment. In the *Apply experiment* step, the scientific solution (experiment) is applied, i.e., the configured experimental apparatus is operated, under conditions defined with given scientific problem parameter values against instances of the phenomenon to produce the scientific solution result that is then analyzed in terms of the hypotheses of the scientific problem to determine the scientific problem result. In practice, this step is completed on the analytical side with the previous Design experiment step to verify the analytical solution prior to mapping it to the scientific side. The previous domain-analysis maps are used to define the final domain-analysis map used to verify the scientific solution (experimental) result and scientific problem result by mapping them to the analytical solution result and analytical problem result. In practice, this step is used to verify what was developed directly on the scientific side. Finally, in the Analyze experiment & result step, information is gathered from the Define, Design, and Apply experiment steps to be analyzed in the Interpret phase to interpret and verify the experiment, and to validate that the scientific problem result are correct both scientifically and analytically.

The *Interpret* phase consists of four steps. In the *Explain solution, interpret result* step, the scientific and analytical solutions and their domain-analysis map are used to explain the solutions, one in terms of the other; and the analytical problem result, and the scientific problem result, and their domain-analysis map are used to interpret the results, one in terms of the other. In the *Verify explanation, validate interpretation* step, the respective explanations are verified, and the respective results are validated using the respective domain-analysis maps and all relevant problem-solving components. In the *Authorize solution & result* step, the verified experiment and the validated experimental results are submitted to the relevant scientific community, e.g., a journal or conference, for authorization as proven scientific results. In the *Curate solution & result* step, the authorized scientific solution and scientific problem result are curated into existing scientific knowledge.

In practice, the scientific problem-solving workflow is applied rigorously following the rules of the scientific method only on the experiment's final, successful execution. The development of an experiment is truly experimental, seldom following the workflow steps as described above and seldom strictly applying the scientific method. Scientists incrementally explore all scientific and analytical problem-solving components to better understand each step[6], as described next for the Higgs boson experiments. The CMS and ATLAS Higgs boson experiments took 48 years of incremental design, development, and validation on both sides, each informing the other.

### 3.2.2. Scientific problem-solving example: the Higgs boson

Consider scientific problem-solving for the scientific problem "does the Higgs boson exist?" (left side of Fig. 4) as conducted in the Higgs boson experiments. The experiments were developed following the scientific workflow by applying the scientific problem-solving paradigm aided by analytical problem-solving that were verified to mirror each other. One aids the other in expressing, deriving, analyzing, discovering, and verifying corresponding models, problems, solutions, and results.

Consider each side separately. The scientific model was the conceptual model of the standard model of particle physics (SMPP) without the Higgs boson. The scientific problem was expressed as an extension of the scientific model in terms of the hypothesized mass of the Higgs boson[7] hence the energy required to move, thus sense, Higgs bosons, and the resulting energy cascades as bosons decay into other elementary particles, e.g., leptons and photons.

The scientific solution (experiment) was the LHC[8] configured using the hypothesized scientific problem parameters that were varied to cover all empirical conditions, i.e., ranges of Higgs boson weights and energy cascades. Operating the configured LHC for the final runs produced the scientific solution (experimental) result, i.e., 2 years of observed mass and energy cascade data. The scientific solution result was analyzed to confirm the hypotheses thus interpreting

---

[6] As Edison said, "I didn't fail 1,000 times. The light bulb was an invention with 1,000 steps."

[7] 125.35 GeV with a precision of 0.15 GeV, an uncertainty of ~0.1%

[8] The LHCis a sophisticated machine designed to directly manipulate and measure properties of individual elementary particles under experimental conditions.

(discovering) the scientific problem result, i.e., the Higgs boson exists with a known mass and energy cascade behavior.

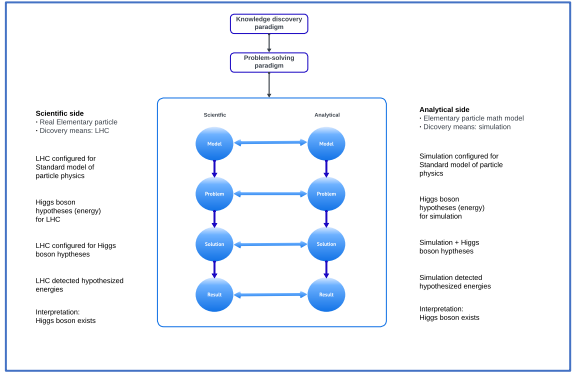

Figure 4: Scientific and analytical problem-solving for the Higgs boson experiments.

Scientific-side discovery was conducted on real physical phenomena, e.g., elementary particles, using the LHC configured for the SMPP, a conceptual scientific model. Analysis-side discovery was conducted using an SMPP simulation (analytical model) with hypotheses (analytical problem) expressed as simulation configurations that varied over conditions analogous to the empirical conditions. The simulation (analytical solution) was conducted over the hypothesized conditions (analytical problem parameter values and data) producing observations (analytical solution result) that were evaluated to produce the simulation result (analytical problem result).

The Higgs boson experiments took 48 years from Peter Higgs' 1964 proposal to the 2012 CMS and Atlas completions. In that time, over 10,000 scientists worldwide explored potential scientific and analytical solutions. As in all scientific experiments, the steps of the Discover and Interpret phases were not sequential but overlapped significantly as did the scientific and analytical problem solutions. Progress was made initially on the analytical side as the SMPP simulation existed and scientific problem-solving was possible only after the LHC completion in 2008. Similar patterns arise in all problem-solving, especially in the formative stages of data science problem-solving, as explored next.

## 4. Data science problem-solving

To better understand data science, the conventional problem-solving paradigm is used to compare the well-understood scientific problem-solving paradigm and workflow with the emerging, inscrutable data science problem-solving paradigm and workflow. The comparison leads to practical insights into the nature of data science and data science problem-solving.

First consider a fundamental difference. In scientific problem-solving, analysis and resulting discoveries are made on the domain side *directly on phenomena*, aided by the analytical side on data representing the phenomena (Fig. 5). *Data science is the reverse*. Analysis and discoveries are made on the analytical side *directly on data*[9] that represents the phenomena.

The complex data science problem-solving paradigm is described below. First data science problem-solving terminology is introduced, followed by a high-level description of the paradigm. Then the role of each data science problem-solving component is explained, followed in §4.2, by their use in the data science problem-solving workflow.

### 4.1. Data science problem-solving paradigm

**Data science terminology** The data science problem-solving paradigm is a specialization of the conventional problem-solving paradigm (Fig. 1). Fig. 6 illustrates a typical application of the data science problem solving paradigm in which only the domain problem is known. Light blue indicates problem-solving components that are often unknown. Mapping analytical terminology to data science terminology[10] requires explanation.

---

[9] Data empiricism – learning from data – versus physical empiricism – learning from phenomena.

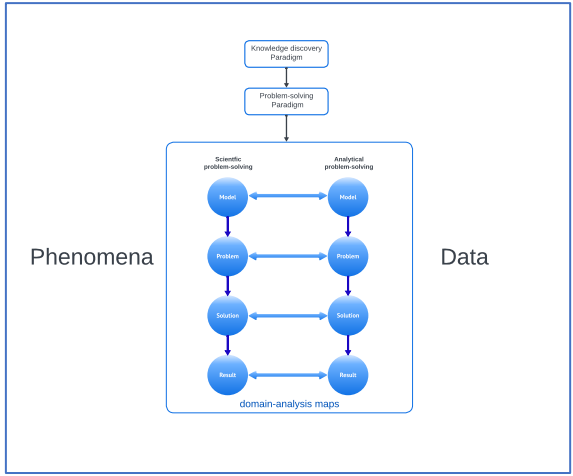

Figure 5: Scientific problem-solving on phenomena.

The conventional problem-solving analytical model and problem correspond to the untrained model and training method, respectively, in data science problem-solving. A data science problem is defined by an untrained model, possibly with *model parameters*, e.g., neural net depth and width. An *untrained model* is a parameterized AI-based data science method that implements a class of analysis[11], i.e., a class of data science problems. The data science problem is further defined by training the untrained model using a *training method*, possibly parameterized by its *data science problem parameters*, applied to *training data*. A *result specification* guides the following steps by qualifying the desired *data science problem result*.

Often, an untrained model is trained using the *training method* and *training data* that is derived from the domain problem and its parameters, intuitively guided by the domain *result specification*. The resulting *trained model* with given data science problem parameter values corresponds to the conventional analytical solution for the parameterized class of data science problems. The trained model can be applied to a specific data science problem instance defined by its *data science problem parameter values* and *inference data*. This produces a *trained model result* that must be evaluated to determine the *data science problem result* that should satisfy the result specification. This terminology emphasizes the unique nature and philosophy of data science problem-solving, distinct from those of conventional and scientific problem-solving, further described in §5.

**Data science problem-solving summary**
The purpose of data science problem-solving for an instance of a domain problem parameterized by its phenomenon parameters is to produce a domain problem result that, to be safe to be applied in practice, satisfies its result specification. The domain problem instance is solved by expressing and solving it as a data science problem instance on the analytical side, as follows.

---

[10] As data science is in its infancy, its concepts and terminology are emerging. To compare data science with science, we use the problem-solving paradigm terminology used in this research.

[11]The large and growing number of data science analysis types are expressed (programmed) by humans often in neural networks including classification, regression, association, anomaly detection, sequence modeling, time series, recommendations, NLP, image recognition, speech recognition, reinforcement learning, etc.

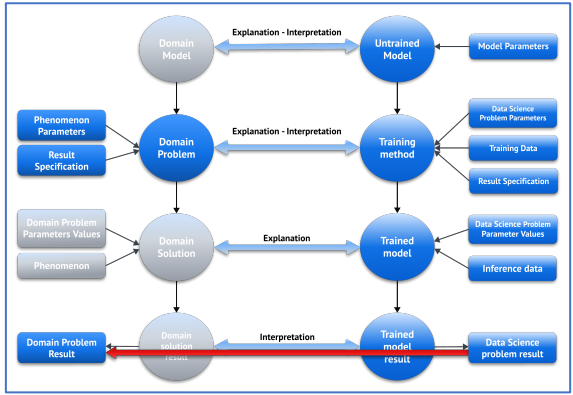

Figure 6: Data science problem-solving paradigm.

An untrained model with model parameters is selected and trained using a training method applied with data science problem parameters and training data to produce a trained model. The trained model is applied with data science problem parameter values and inference data that represents the problem instance producing a trained model result that must be interpreted to produce a data science problem result that must satisfy its result specification. The data science problem result must be interpreted in terms of the desired domain problem result on the domain side by a human-created domain-analysis map (in red in Fig. 6). Since data science cannot be used to produce or verify either the domain problem result nor the domain-analysis map, means outside data science must be used to ensure that it satisfies its result specification.

**Data science problem-solving components**
A domain problem expresses knowledge to be discovered or questions to be answered about instances of a phenomenon. Often, a domain problem is known, as are phenomenon parameters that characterize the knowledge being sought. A domain problem can be used to gain insights into its domain model, i.e., a model of the phenomenon being analyzed, hence a context for the domain problem. Alternatively, the domain model of a phenomenon is known of which one or more domain problems are to be analyzed. A phenomenon can be analyzed from many perspectives, each expressed as a distinct domain problem. A result specification is used to characterize the domain problem result to aid in defining and verifying a domain problem result and the data science problem-solving components used to produce it.

The domain model, domain problem, phenomenon parameters, and the result specification are used to gain insights into creating or selecting an untrained model, its training method using or resulting in human created domain-analysis maps, i.e., what type of analysis will solve the problem? An untrained model is selected from a library or implemented using an untrained AI-based data science method, often in a neural network, that conducts the intended analysis type to solve a data science problem corresponding to the domain problem. The training method and other components – domain problem and its phenomenon parameters, domain model, untrained model, and any domain-analysis maps – are used to define training data[12] and inference data including their data quality requirements and acquisition methods. The untrained model is trained using the training method applied to training data to produce a trained model with specifications for inference data and for data science problem parameter values, i.e., parameters derived from the domain problem and data science model tuning hyperparameters. The trained model is tuned, using data science problem parameter values, i.e., hyperparameter values, and is tested over critical requirements using test inference data. As most trained AI-based models cannot be validated to meet specific requirements, e.g., accuracy and safety, guardrails[13] are often required to limit its behavior, i.e., trained model result, within bounds defined by a result specification. The resulting tested, trained model can be applied to a data science problem instance defined by its data science problem parameter values and inference data. This produces a trained model result that is interpreted as a data science problem result for the data science problem instance. The result specification is used to ensure that the data science problem result is within bounds. The motivating domain problem instance is solved only when 1) the data science problem result has been interpreted in terms of the desired domain problem result, with a plausible domain-analysis map[14], and 2) means outside of data science have been used to verify that the domain problem result satisfies its result specification. A simple domain-analysis map example is in convolutional neural network image recognition in which supervised learning establishes a map between generic images, e.g., of bananas, and the label "banana". The trained model knows nothing of such interpretations, e.g., of bananas nor of English terms used to label bananas; it programmatically maps the label "banana" to images that it recognized as bananas.

---

[12] This valuable insight leads to means for defining and acquiring the most critical element, high quality data and its defining requirements and embeddings. Not all training methods require training data.

Data science problem-solving requires that all analytical problem-solving components (Fig. 6 right side) be completed. Often, the domain problem is the only domain side problem-solving component that is known. Exploratory data science starts without a domain problem that is discovered in data that represents a phenomenon. A domain problem can be known without knowing its domain model since the purpose of a domain problem is to discover the nature of the phenomenon being analyzed. While it seems odd to solve a problem without understanding its context, data science enables such exploration due to its scope, scale, complexity, and power to learn from data [3][4]. *Data science enables such previously impossible knowledge discovery.*

All data science workflow phases and steps are aided using domain-analysis maps and derivation relationships. Due to the inscrutability of AI-based data science, they too are inscrutable. Attempting to define them can provide valuable insights into understanding, developing, and verifying related components. If the results are to be applied in practice, two domain-analysis maps are critical and necessarily require using means outside data science to establish and verify them. First, to apply a domain problem result in practice, it must satisfy the domain problem result specification; insights for this can be gained by ensuring that the corresponding data science problem result satisfies the data science result specification. Second, it is essential to establish a correspondence between the motivating domain problem and the training method to verify or demonstrate, if only intuitively, that the data science problem being addressed corresponds to the motivating domain problem, i.e., was the analysis conducted of the intended type? Often, the domain solution is unknown. If a domain problem result can be demonstrated outside of data science to satisfy its result specification, a domain solution is not required, and may be beyond human understanding in scale and complexity, as in AlphaGo and AlphaFold [19].

---

[13] Guardrails are empirically discovered and evaluated constraints on trained models intended to bound behavior, but without certainty. There is considerable research in developing these practical necessities.

[14] The domain-analysis maps back to the domain side are called *explanation* for the trained model result and *interpretation* for the data science problem result. They map or *de-embed* data science data to domain data.

### 4.2. The data science problem-solving workflow

As with all problem-solving, data science problem-solving has myriad workflows depending on what problem-solving components are known. The following is a typical data science problem-solving workflow (Fig. 7) corresponding to the typical data science problem in Fig 6. Many more can be deduced from insights in §5.

Data science problem-solving is conducted following the data science method in the two phase data problem-solving workflow (Fig. 7) using data science problem-solving paradigm components (Fig. 6) as follows and demonstrated with AlphaFold in §4.3.

The *Discover* phase consists of four steps. In the *Define* step, a motivating domain problem with its domain problem parameters is defined, i.e., a specific question or analysis of a phenomenon is defined, within a relevant domain model. As explained above, a domain model of the phenomenon being analyzed may not be adequately known as the domain problem may be learned from data.

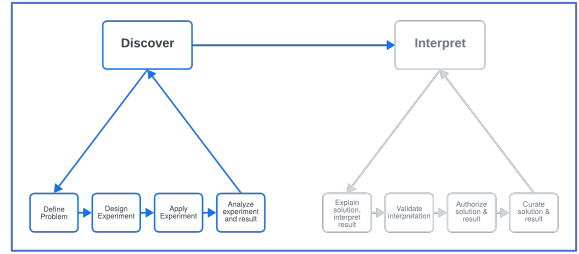

Figure 7: Data science problem-solving workflow

The Define step provides insights into the *Design* step that is used to design or select from a library an untrained model, i.e., a data science analytical method that implements the class of analysis required to answer the domain problem, together with its training method. The training method and other components – domain problem and its phenomenon parameters, domain model, untrained model, and any domain-analysis maps – are used to define training data and inference data including their data quality requirements, acquisition method, and embeddings. The untrained model is trained by applying the training method to the training data producing a trained model with specifications for its data science problem parameters and inference data that may be derived from knowledge of the untrained model, training method, and training data and insights from the domain side. The resulting trained model is tuned by adjusting the data science parameter values, i.e., hyperparameters, and testing the trained model with test inference data. Guardrails are developed to attempt, without guarantee, to limit the trained model results, i.e., the behavior of the trained model, to satisfy the result specification.

In the *Apply* step, the trained model is applied to a specific data science problem instance defined by its inference data and data science parameter values producing a trained model result that must be evaluated to produce a data science problem result which should satisfy the result specification. No results – trained model result, data science problem result, hence domain problem result, can be relied upon to satisfy result specification when applied in practice even if they do so in developing the results. Result specifications are metrics used to guide, without guarantee, the development of a trained model, i.e., data science solution, to be applied to data science problem instances that are safe to apply in practice. Finally, in the *Analyze* step, all problem-solving components and their results are analyzed in preparation for the *Interpret* phase.

Due to the inscrutability of AI-based data science, there is no theoretical or practical basis, as there is in science, for the Interpret phase. The four step *Interpret* phase must be fabricated by humans based on intuition, experience, and expertise with empirically developed tools, i.e., guardrails, and methods outside the field of data science. The *Explain & interpret* step attempts to 1) explain the analysis conducted by the trained model, ideally but rarely confirming that it corresponded to the intended domain solution, and 2) interpret the trained model result in terms of the data science problem result. In simple cases, the two results are identical. For complex trained models, such as the AlphaFold [19] ensemble model, the trained model result must be interpreted in terms of the data science problem result. The Explain & interpret step interprets the data science problem result in terms of the motivating domain problem result. The Validate & verify step can be very complex requiring considerable human intuition and domain expertise. The *Authorize* and *Curate* steps are like those in scientific problem-solving. The analysis and its results are documented and submitted for authorization, in research to a journal or conference, and in industry to a relevant institutional authority. While methods for proving, hence authorizing, scientific results are well-known, means of consistently demonstrating, not even proving, properties of data science results are just emerging.

In practice, data science problem-solving ranges from simple, developed and applied in hours, to complex taking a decade to develop, as in the case of AlphaFold, described next. Simple cases involve existing trained models in AI applications, e.g., OpenAI's ChatGPT, simply require input prompts and parameters without modifying the underlying trained model. However, trained models reflect only the knowledge on which they were trained, hence do not reflect subsequently developed knowledge. Existing trained models, especially Foundation models, can be updated or refined by repeating the Discover phase.

### 4.3. Data science problem solving example: AlphaFold

The above data science problem-solving paradigm and workflow descriptions are for simple data science problems with a single data science analytical method and model. Data science problems and solutions can be considerably more complex, requiring many types of expertise and models and considerable time to develop and train. For example, consider DeepMind's AlphaFold[15] [19], one of the most successful and complex AI-based data science solutions

---

[15] Each of DeepMind's series of Alpha models was an unprecedented milestone in the application of AI-based data science to address complex, previously unsolved problems, each surpassing the former.

(illustrated in Fig. 8, light blue indicates components that are mostly unknown).

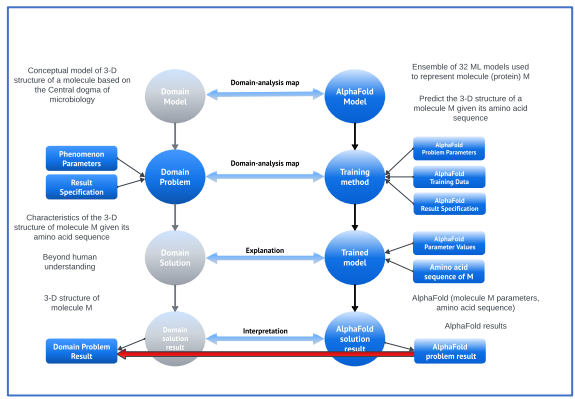

Figure 8: Data science problem-solving for AlphaFold

AlphaFold is a revolutionary technology and breakthrough success that accurately solves a 50 year old microbiology grand challenge – predicting the 3D structure of a molecule given its amino acid sequence. Domain side problem-solving components are at best partially known, e.g., there may not be a scientifically verified domain model, i.e., the conceptual model for the 3D structure of all molecules. Prior to AlphaFold, over 30 years of experimentation produced costly, imprecise computational analytical models and solutions that could take up to ten years for one molecule. The partially understood domain model and well-understood domain problem may have provided insights to develop AlphaFold model – a complex ensemble of 32 deep learning models. The domain solution and domain solution result are shaded light blue in Fig. 8 to suggest that they are not known. But they are not required as AlphaFold produces the desired domain problem result directly from the AlphaFold domain problem result, developed with human knowledge and insight. The AlphaFold solution applied with inputs, the AlphaFold parameter values and Amino acid sequence of a molecule, produces the AlphaFold solution result that is then interpreted to produce the AlphaFold problem result that is further interpreted (via the red domain-analysis map in Fig. 8) to produce the desired domain problem result. The AlphaFold solution and result, and interpretation domain-analysis map are inscrutable within data science, i.e., lack full explanations and robust interpretations – neither understood nor provable. Their validity is established with means outside of data science such as human knowledge, expertise, experience, and scientific knowledge, e.g., cryo-electron microscopy.

AlphaFold illustrates the nature, complexity, and scale of problems that can be successfully addressed with AI-based data science. While the exact size and composition of the AlphaFold team and their expertise is not publicly available, it has been estimated [39] to be 50-100 scientists with world-class expertise in protein structure biology, machine learning, computational biology, software engineering, structural bioinformatics, biophysics, and mathematics.

## 5. Scientific versus data science problem-solving

This section summarizes critical differences between science and data science, introduced above and in [3][4]. The scientific and data science problem-solving paradigms are specializations of the conventional problem-solving paradigm; yet they differ fundamentally in nature and detail. Hence, the conventional problem-solving paradigm is a framework with which to compare the two paradigms, to understand their similarities and differences, and to gain insights into data science and data science problem-solving. *Scientific and data science problem-solving take different paths to fundamentally different solutions.* This and subsequent italicized speculations relate to Stephen Hawking's final theory [13].

### 5.1. Results in science are universal, in data science local and specific

A fundamental difference between scientific and data science problem-solving, stated simply, is that while both are evaluated on specific instances of a phenomenon, scientific results are claimed to be certain and apply universally to all such phenomena under the same empirical conditions; data science results are interpreted as uncertain insights into an instance or class of instances of a phenomenon.

Following the scientific method, a scientific solution (experiment) is successful if it is based on verified scientific knowledge, the experiment and results are reproducible, all of which has been documented, verified, and authorized by the relevant scientific community. The resulting scientific knowledge is fully understood and is considered, by scientific induction[16], to be universally true under the empirical conditions, thus definitive. Such experiments are conducted twice – once to establish the scientific problem result (i.e., scientific knowledge) and once to prove repeatability; after which the experiment need never be repeated.

Following the data science method, a successful data science solution (trained model) is applied to a specific instance, or class of instances, of a data science problem defined by its inference data and data science problem parameter values. Data science is in its infancy and is evolving rapidly, hence, the following observations may also evolve; just as science did when it emerged centuries ago and does to this day. First, AI-based data science solutions (trained models) are inscrutable and produce results that are uncertain, i.e., not definitive, and cannot be proven true within data science. At best, data science problem results are local, specific to a trained model instance, its untrained model, training method and training data, and to the data science problem instance, defined by its data science problem parameter values and inference data. Unlike a universal scientific solution, a data science solution is intended to be applied repeatedly, each time to a different data science problem instance. As a data science solution and its data science problem parameter values and inference data are developed using a limited training method and limited training data, it is common practice for each problem instance to update the solution with additional training and tuning. Also, in contrast with universal scientific solutions, data science solutions are not only local and specific to their development, they are always imperfect, hence subject to revision to improve the data science problem result for a motivating domain problem and domain problem instance. This leads to an interesting tradeoff. Like most inscrutable properties of data science solutions, their locality versus universality and their quality can often be improved. In contrast with scientific results that are definitive; data science results need only be *good enough,* i.e., satisfy the result specification for the motivating domain problem. At the same time, the unbounded potential of data science solutions for reuse and revision reflect the corresponding unfathomed scope and power of data science problem-solving. In contrast with the scientific method, the data science method supports a range of solutions for the same motivating domain problem that can vary depending on their development — purpose of the motivating domain problem, training method, training data, parameters, and inference data. *Local data science results may be more realistic hence more informative and valuable than alleged universal scientific knowledge [4].*

---

[16] Scientific induction has been disputed since 1748 by Hume [16][17]

### 5.2. Problems and solutions in science are human-scale, in data science unfathomed

For conceptual and practical reasons, scientific models and results are at human scale, e.g., fully understandable by a scientist aided by models, e.g., natural language, mathematical, or simulation models. Following the scientific method, an experiment must evaluate all combinations of the factors that potentially cause a hypothesized effect. To fully evaluate N potentially casual factors requires that all combinations of those factors, i.e., $2^N$ cases, be evaluated. For practical reasons, most experiments explore less than 10 factors requiring 1,024 cases and more typically 4 factors requiring 16 cases. In contrast, data science problems and results and their corresponding domain problem results can vastly exceed human scale, e.g., Nvidia's LLM Megatron-Turing NLG 2.0 has 1 trillion parameters. The scale, i.e., complexity, of problems and solutions that can be addressed by AI-based data science is unfathomed. *Data science offers insights into instances of complex problems not previously nor otherwise achievable*.

### 5.3. Solutions are complete in science, incomplete in data science

Scientific problem-solving is complete with all scientific and analytical problem-solving components fully understood, as are domain-analysis maps used to prove equivalence and derivation relationships used to derive components. Similarly, all steps of the Discover and Interpret phases of the scientific problem-solving workflow are known with proven results. Hence, scientific knowledge discovery is fully understood with the successful application of the scientific method producing proven, i.e., certain, scientific knowledge.

In contrast, data science problem-solving is incomplete with problem solving components unknown to varying degrees (light blue shading in Fig. 6-9). Consider the domain side. Often, the domain problem is known, but may have resulted from a previous data analysis. A domain solution is rarely known and may be beyond human understanding; however, it is not needed if an adequate domain problem result has been established. While domain-analysis maps and derivation relationships are inscrutable, intuiting such maps aids defining analytical components. A domain problem can suggest the class of analysis to be conducted, hence an appropriate untrained model with its training method and training data. With so little known on the domain side, analytical components must be developed with little guidance, save intuition, experience, and knowledge, e.g., pre-trained models such as Foundation models. In contrast with science's tractable Interpret phase, defined by the scientific method, the data science method lacks a tractable Interpret phase. It must be fabricated based on intuition, knowledge, and expertise with empirically developed tools, i.e., guardrails, and methods outside data science as no theoretically-based methods exist.

### 5.4. Knowledge discovered in science is certain, in data science is uncertain.

Scientific results are authorized as true by being accepted by the relevant expert community. Scientific approvals are based on centuries of applying the scientific method resulting in the certainty of scientific knowledge, hence the certainty of our knowledge of the natural world. This contrasts with the uncertainty of data science solutions and results. "Knowledge" discovered by data science is inscrutable, probabilistic, possibly erroneous, lacking theoretical or practical means for explanations and interpretations. To be applied in practice, means outside data science, especially empirically developed guardrails, are used to limit resulting domain problem results within previously specified bounds. However, as quantum mechanics demonstrated 100 years ago[17], reality may not be as certain as assumed in science but probabilistically uncertain as in data science. *The uncertainty and incompleteness in the emerging philosophy of data science may be more realistic than certainty and completeness in science, possibly leading to transforming and redefining science [4].*

### 5.5. Science analyzes real things; data science real and abstract things

In scientific problem-solving, discoveries are made on the scientific (domain) side aided by analysis on the analytical side (Fig. 9, 10). Following the scientific method, phenomena are central as knowledge is discovered directly from phenomena instances by well-understood *physical empiricism* and scientific theories, producing certain scientific results. In science, data on the analytical side, is secondary, used to develop an analytical solution and result to guide or to confirm experiments via domain-analysis maps. The scope of science is limited to real, measurable phenomena at human scale.

Data science problem-solving is the reverse. Discoveries are made (learned) from data on the analytical side possibly aided by intuition for interpretations from the domain side. Data is central since *for knowledge to be discovered, it must be in the data.* The scope of data science is any real, abstract, or imaginary, phenomenon for which there is adequate data, potentially at scale beyond human understanding. Hence, data science is data-centric[18] (data intensive, data-driven). Discoveries are made directly from data following the data science method by inscrutable *data empiricism* with uncertain results. AI-based data science analytical methods are secondary as there are many with which to discover knowledge in data. *Unlike 20th C data that are assets, 21st C data science data is phenomenological – a resource in which to discover phenomena and their properties, previously and otherwise impossible [5].*

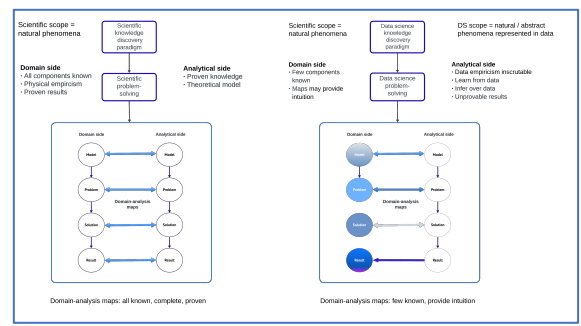

Figure 9: Scientific versus data science problem-solving components.

### 5.6. Results derived by humans in science, by computation in data science

Scientific problem-solving is complete with all problem solving components – models,

---

[17] Einstein, by his own admission, was not comfortable with, educated or practiced in reasoning in uncertainty with uncertain, probabilistic, ambiguous outcomes, despite increasing evidence in modern physics.

[18] Due to the power of AI-based data science methods, data science was initially considered model-centric.

problems, solutions, and results, both scientific and analytic, domain-analysis maps, and derivations – developed, fully understood, and proven by humans. In data science, only the untrained model problem-solving component, often expressed as a neural network, is developed and largely understood by humans. Even the domain problem may be initially unknown, discovered computationally. All other problem-solving components – training method, trained model, trained model result, data science problem result and even the domain problem result are developed computationally and are inscrutable. Consider the nature of what is developed computationally. An AI-based analytical model – an untrained model – is often designed by humans in a neural network hence is largely understood. Yet, the "knowledge" that it contains, i.e., the class of analysis to be conducted such as image recognition, can be inscrutable. A training method is used to train an untrained model with training data that specializes the initially human-defined data science analysis, i.e., untrained model, to solve the problem in specific instances, e.g., what properties uniquely identify pizzas and teapots, to produce a trained model. It can computationally establish a domain-analysis map, e.g., properties identifiable by the trained model of all images of pizzas in inference data with those properties as a "pizza". How a training method trains an untrained model, e.g., what properties define a pizza or a teapot, or how an untrained model learns from the training is inscrutable. Thus, as if by magic, a trained model with data science problem parameters values (e.g., hyperparameters) is applied to previously unseen inference data, e.g., a pizza image, and solves that problem instance producing a trained model result, i.e., it has the properties of a pizza that is (in this case trivially) interpreted as the data science problem result that it interprets by applying the domain-analysis map to produce the desired domain problem result, i.e., "It is an image of a pizza". The training, learning, inference is at the heart of the powerful but inscrutable data science knowledge discovery paradigm.

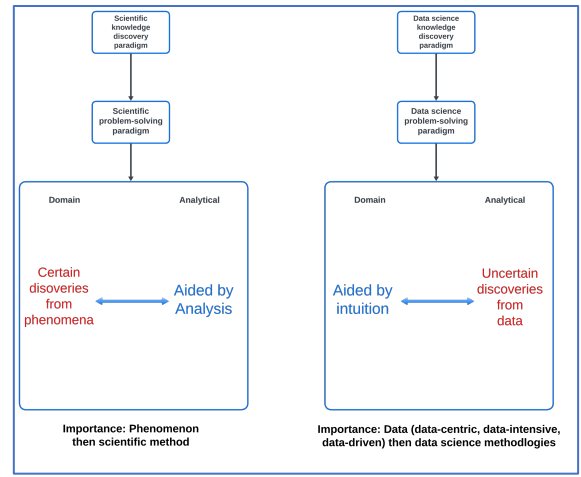

Figure 10: Scientific versus data science problem-solving nature.

### 5.7. *Value: understood, limited in science; inscrutable, unbounded in data science*

Consider the above fascinating differences further. The value of science, for centuries our most powerful knowledge discovery paradigm, is as a method of discovering scientific knowledge of natural phenomena of our universe. Science and the resulting scientific knowledge are fully understood but limited to measurable real phenomena. The value of data science will emerge as an even more powerful knowledge discovery paradigm, despite its inscrutability. Data is emerging as a source of knowledge never before imagined, yet to be understood [5].

Consider the inscrutability of data science methods and data as compelling mysteries at the heart of AI-based data science. The extent to which a trained model, i.e., a data science solution, can discover valuable patterns of values of properties of a phenomenon, i.e., a desired data science solution result, in previously unseen inference data depends on at least three things. First and foremost, the pattern must be in the inference data. This could be verified outside data science by a costly, exhaustive analysis. Second, the untrained model, i.e., analytical method, underlying the trained model must be capable of discovering the pattern. This cannot be verified and is the subject of exhaustive empirical testing at great expense. Third, the ability to discover the pattern must have been learned from the training data, i.e., must have been present in the training data sufficiently for the pattern to be incorporated in the trained model. This too cannot be proven and is being explored empirically and otherwise. While this inscrutability poses significant challenges for validating data science problem-solving, understanding them may lead to solution insights and opportunities summarized in §6. Finally, the power of data science transforms the role, e.g., conception, of data from 17th C *objects* used to record facts and 20th C *assets* used to manage information [21][22] to *phenomenological* in the 21st C – a source for knowledge discovery of any phenomena for which there is adequate data [5].

## 6. Insights from data science problem-solving scenarios

Despite the framework being graphically simple, it can be used to understand the inherently complex data science problem-solving paradigm and solutions. Each component (bubble, rectangle) and arrow (domain-analysis map, derivation) has a significant meaning and role in data science problem-solving. These aid understanding as each component can provide insights into data-science problem-solving problems, solutions, and results. Due to the immaturity and rapid development of the data science problem-solving paradigm, insights are required to understand, develop, produce, and verify prospective models, problems, solutions, results, domain-analysis maps, derivations, workflows, and the continuous process of testing and refining them and interpreting them in terms of motivating domain problems and desired domain problem results.

This section summaries such insights introduced in earlier sections. Most such insights come from the subjects of analyses – real or abstract phenomena on the domain side – into the inscrutable data science solutions and results on the analytical side, and *vice versa*.

### 6.1. Domain Model

Domain problems can be understood as hypotheses within a domain model or context. For a given domain problem, the domain model may be known, or unknown but intuited. It may provide intuition for, or insights into, the domain problem, and, with similarly human-intuited domain-analysis maps, one or more untrained models, i.e., analytical categories. For example, a domain problem concerning a domain model, i.e., of a phenomenon, could be analyzed in multiple ways, i.e., as multiple data science problems, each with its own training method, parameters, training data, and result specification. A domain model may be simple, representing a single, simple phenomenon, or could be complex, composed of many phenomena, e.g., an amino acid sequence of a protein. A domain model should provide intuitive insights into the domain problem, and domain problem result and *vice versa*. The result of such analyses may be a combination of the solutions as in an ensemble model. A domain problem is often solved using data science problem-solving without fully defining the domain model or context. Due to the potentially complex nature of data science trained models and problem results, and the corresponding domain problem result, there may not have been an initial domain model. A common use of data science problem solving is to learn from and discover in data, properties of phenomena. Multiple such discoveries can contribute to defining or elaborating a domain model.

### 6.2. Domain model, problem, phenomenon parameters, and result specification

Data science problems often start with a domain problem to be solved with an intuitive notion of a result specification and of the phenomenon parameters that characterize the knowledge to be discovered, i.e., a domain problem result – the properties, or patterns of properties of the phenomenon to be analyzed. Data science problem-solving is often used when there is no domain solution[19] with which to directly produce a domain problem result, but there is adequate data representing instances of the properties of the phenomenon to be analyzed and one or more untrained models, i.e., AI-based data science analytical methods, to produce the desired domain problem result or provide insights into finding such a result.

Domain problems can range from being well-defined, to ill-defined, to unknown. The extent to which a domain problem is defined provides intuition for or insights into the following, each aided by an intuitive but inscrutable domain-analysis map or derivation relationship. All problem-solving components are selected by humans based on intuition[20] – possibly gained from domain side components – knowledge, experience, and previously developed solutions.

- If the domain problem is unknown, as in exploratory AI, a human can explore an untrained or trained model to analyze candidate inference data to find patterns to be used as hypotheses to define a domain problem.
- Domain model – a human-defined generalization of the domain problem, i.e., properties, or patterns of properties, of the phenomenon to be analyzed.
- Untrained model – an untrained, AI-based method, selected by a human by intuition, experience or knowledge to conduct the intended category of analysis specialized by its training method, data science problem parameters, and training data that is required to discover the desired domain problem result for a domain problem instance that satisfies its result specification.
- Training method – an AI-based method for using training data to train an

---

[19] This is an example of *data science thinking* – producing a possibly uncertain domain problem result from data in the absence of a domain solution. In *scientific thinking*, proven scientific knowledge (results) are established by means of a verified scientific solution.

[20] AI-based data science has captured the world's attention due to its problem-solving scope, scale, complexity, and power previously impossible with human coded computations. Yet it is inscrutable. The *Holy Grail of AI-based data science* is to explain how it works and to interpret results in terms of the motivating domain problem. While this is the case, possibly forever, human intuition, guidance, and reasoning are critical for AI-based data science solutions to be safe and *good enough* in practice.

- untrained model to discover an desired domain problem result in previously unseen inference data.
- Training data – data that represents knowledge of the properties, or of patterns of properties of instances of the phenomenon to be discovered by the trained model in previously unseen inference data.
- Result specification – a characterization of the intended data science problem result that must be interpreted in terms of the domain problem result that must satisfy the domain problem result specification.
- Trained model – a trained AI-based method parameterized by data science problem parameter values that together with the inference data define the instances of the phenomenon to be analyzed. In rare cases, human intuition might be used to develop a domain-analysis map to an intuited domain solution.
- Trained model result – the result of applying the trained model with specific data science problem parameter values to inference data. It may be possible to intuitively verify the correctness of the computational result of the trained model application.
- Data science problem result – the interpretation of the trained model result in terms of the data science problem. Considering the data science problem as a hypothesis, does the trained model result confirm or deny that hypothesis?
- Domain problem result – the desired result for a specific instance of the domain problem. It should meet its domain problem result specifications. It must be interpreted entirely by human intuition, experience, and knowledge as there are no means within data science to do so. This critical interpretation is made primarily based on the data science problem result that in turn is based on the data science problem solving workflow steps and components that produced the data science problem result.

Finally, the quality of a data science problem result and of the desired domain problem result depends directly on data quality [5], i.e., requirements of data – the most critical component in data science analyses. The corresponding most valuable insights are to be gained from the domain problem, phenomenon parameters, and result specification components that concern the training method, training data, result specification, and inference data. Training data determines what the untrained method will learn. Inference data determines what the trained model will infer, to produce the data science problem result that leads to the domain problem result. All domain side components provide insights into identifying, acquiring, evaluating, and refining the required data.

### 6.3. Domain solution, domain problem parameter values and the phenomenon

Frequently, a domain solution is unknown or is beyond human understanding in complexity. This motivates the use of data science to find a data science solution, i.e., a trained model, to be applied to specific instances of the domain problem. Trained models are inscrutable hence cannot be proven to be correct but may be demonstrated by means outside of data science to produce data science and domain problem results that satisfy data science and domain result specifications. Such means are developed intuitively using relevant expertise, experience, and empiricism as described for AlphaFold in §4.3. Such demonstrations can be developed only for specific domain problem instances and corresponding data science problem instances, not for all such instances. While a domain problem result demonstrated to satisfy its result specification obviates the need for a domain solution, the contributing problem-solving components may provide insights into the motivating domain problem instance, the domain problem, the domain model, and ultimately into the domain solution. While many models contribute to understanding protein folding (hydrophobic collapse; lattice, framework, and coarse-grained models), there is currently no domain model or solution. Some problems, like protein folding, may be too complex to admit of a domain solution for domain problem instances in contrast with the universality of scientific solutions.

### 6.4. Training and inference data; untrained models and training methods

The most important data science problem-solving components are first, training and inference data, and second, an untrained model and its training method. Data is most important since *for knowledge to be discovered, it must be in the data.* While there may be many insights into the nature of the data, e.g., data typically used to describe a phenomenon, such conventional data may exclude knowledge being sought. Solving such problems required data thinking [13]. Consider knowledge of astrophysical phenomena that emerged and possibly vanished since the origins of time. The James Web Space Telescope records data from the beginning of time - 13.6 B ($10^{11}$) years ago. Training and inference data for data science analyses is represented with conventional data structures[21]. What training and inference data is required to discover previously unknown astrophysical phenomena? As with the AlphaFold description in §4.3, there is no known theory or practice to define that data *a priori.* Are conventional data structures adequate or do they preclude that very knowledge being sought? Insights must be gained by trial and error based on intuition, experience, and knowledge from many disciplines exploiting the data science problem-solving components as suggested above.

An untrained model and its training method are the next most important problem-solving components. Just as critical data is discovered by insights through trial and error, so too are untrained models, i.e., categories of analysis required to discover patterns of those phenomena, and training method and training data required to learn the patterns to produce a trained model to enable such discoveries for specific instances of those phenomena, represented in the previously unseen inference data. Insights will come from trial and error evaluating many untrained models, i.e., categories of analysis . Insights may come from proven data science analysis results for known phenomena, or may require novel AI-based analytical methods, i.e., untrained models.

AI-based data science is in its infancy, with its greatest successes based on neural networks – the solutions referenced in this paper. Neural networks have limitations and are not universal problem-solving architectures. Many new problem solving architectures are emerging thus expanding AI-based data science. "The future of machine learning architectures promises exciting developments that will lead to more powerful and versatile AI systems capable of tackling even more complex challenges."[22]

---

[21] Infrared electromagnetic radiation (EMR) data stored as common extensible Markup Language (XML) in the command and telemetry database [11].

## 7. Conclusion

This paper provides a framework with which to better understand the data science problem-solving paradigm and workflow and illustrates many resulting insights.

*As data science is better understood, it will our most powerful source of insights into our inherently uncertain, probabilistic world. Its dominant contribution may be its scope, scale, complexity, and power for gaining insights into our universe.*

## 8. Appendices

### 8.1. Many useful definitions.

Over 200 years scientists have developed many useful definitions of *the scientific method* and its expression in *scientific problem-solving paradigms* and *scientific workflows* each accepted by the relevant scientific communities. No single definition satisfies all requirements. This paper uses a simple, incomplete definitions selected to enable comparison with the emerging and inherently complex data science. Similarly, many definitions of *the data science method* and its expression in *data science problem-solving paradigms* and *data science workflows* are emerging reflecting the disciplines and problem classes that they serve. Due to the rapid development of data science, they will continue to evolve. Those offered here are intended to explore the inscrutable yet powerful data science problem-solving paradigm and workflow at this, the beginning of a decades-long discovery process.

### 8.2. Data science paradigm: learning from data

The framework presented here applies to all of data science but addresses AI-based data science that, due to its inscrutability, poses the greatest challenges. It does not address conventional data science, as defined below, excerpted from [3].

"The *philosophy of data science* is the worldview that provides the philosophical underpinnings (i.e., learning from data) for data science research for knowledge discovery with which to reason about (understand), discover, articulate, and validate *insights into the true nature of the ultimate questions about a phenomenon by computational analyses of a dataset that represents features of interest of some subset of the population of the phenomenon.* Data science results are probabilistic, correlational, possibly fragile or specific to the analysis method or dataset, cannot be proven complete or correct, and lack explanations and interpretations for the motivating domain problem."

While the term data science is new, the field of data science is as old as mathematics, our most widely used method for learning from data. Pre-AI data science, aka conventional data science, includes methods such as mathematics, simulation, databases, data mining, statistics, probability theory, approximation theory, and some AI techniques like decision trees and linear SVMs. While conventional data science methods have been used for centuries, only now are they recognized as the only transparent – scrutable – methods and, now, the least powerful. In this well understood category, solution explanations and results interpretations, while not inherent, are easier to construct than for AI-based methods, e.g., weather prediction models are designed, explained, and interpreted by experts using complex mathematics and simulations. Conventional methods and models are designed by humans to meet specific requirements hence humans are the agents of learning. AI-based methods emerged in the 1990s and now dominate data science. While designed by humans, they are developed computationally by AI algorithms such as machine learning (ML), evolutionary, heuristic, and generative algorithms. AI-based methods are inscrutable lacking solution explanations and results interpretations. In conventional data science methods, humans are the learning agents. In AI-based data science methods, algorithms are the learning agents. That difference alone – human versus algorithmic learning – distinguishes conventional versus AI-based data science. It also leads to the inscrutable scope, scale, complexity and power of AI-based data science.

---

[22] Conclusion of a response from Google's Gemini (2.12.24) to the prompt "Please identify emerging AI-based data science problem-solving architectures." Gemini identified ten classes of such architectures from verified published research papers.

**Acknowledgement**
I am grateful to Daniel Fischer, Sr. Data Scientist in J.P. Morgan's A.I. Research Division for valuable contributions to this work.

9. References

1. Bento M, Fantini I, Park J, Rittner L, Frayne R. Deep Learning in Large and Multi-Site Structural Brain MR Imaging Datasets. Front Neuroinform. 2022 Jan 20;15:805669. doi: 10.3389/fninf.2021.805669. PMID: 35126080; PMCID: PMC8811356.

2. Britain's NHS is trying once again to collate patients' data: The project is imperfect and controversial, but the technology is needed, The Economist, Oct 18th, 2023.

3. Brodie, M.L., Defining data science: a new field of inquiry, arXiv preprint https://doi.org/10.48550/arXiv.2306.16177 Harvard University, July 2023.

4. Brodie, M.L., A data science axiology: the nature, value, and risks of data science, arXiv preprint http://arxiv.org/abs/2307.10460 Harvard University, July 2023.

5. Brodie, M.L., Re-conceiving data in the 21st Century. Work in progress, Harvard University.

6. Casey BJ et al. The Adolescent Brain Cognitive Development (ABCD) study: imaging acquisition across 21 sites. Dev. Cogn. Neurosci. 32, 43–54 (2018). [PubMed: 29567376]

7. Densen P. Challenges and opportunities facing medical education. Trans Am Clin Climatol Assoc. 2011;122:48-58. PMID: 21686208; PMCID: PMC3116346.

8. Donoho, David (2017) 50 Years of Data Science, Journal of Computational and GraphicalStatistics, 26:4, 745-766, DOI: 10.1080/10618600.2017.1384734 (republished with comments)

9. Eubanks, Virginia, 1972-, Automating Inequality: How High-tech Tools Profile, Police, and Punish the Poor. New York, NY, St. Martin's Press, 2018.

10. Franks, Bill. 97 Things About Ethics Everyone in Data Science Should Know. United States: O'Reilly Media, 2020.

11. Gal-Edd, J. and Fatig, C.C., 2006, March. James Webb Space Telescope XML

database: from the beginning to today. In 2006 IEEE Aerospace Conference (pp. 7-pp). IEEE.

12. Hacking, I. (2006). The Emergence of Probability: A Philosophical Study of Early Ideas about Probability, Induction and Statistical Inference (2nd ed.). Cambridge: Cambridge University Press. doi:10.1017/CBO9780511817557

13. Hazzan, O., Mike, K. (2023). Guide to Teaching Data Science: An Interdisciplinary Approach. Springer International Publishing, 2023. https://doi.org/10.1007/978-3-031- 24758-3_1

14. Hertog, Thomas. On the Origin of Time: Stephen Hawking's Final Theory. Bantam, 2023.

15. Horien C, Noble S, Greene AS, Lee K, Barron DS, Gao S, O'Connor D, Salehi M, Dadashkarimi J, Shen X, Lake EMR, Constable RT, Scheinost D. A hitchhiker's guide to working with large, open-source neuroimaging datasets. Nat Hum Behav. 2021 Feb;5(2):185-193. doi: 10.1038/s41562-020-01005-4. Epub 2020 Dec 7. PMID: 33288916; PMCID: PMC7992920.

16. How to make Britain's health service AI-ready: The NHS should clean up and open up its data. Patients will benefit. Oct 19th, 2023.

17. Hume, David. An Enquiry Concerning Human Understanding, Oxford: Oxford University Press, 1748

18. Hume, David. An Enquiry Concerning Human Understanding. United Kingdom, Oxford University Press, 1999.

19. Jumper, J., Evans, R., Pritzel, A. et al. Highly accurate protein structure prediction with AlphaFold. Nature 596, 583–589 (2021). https://doi.org/10.1038/s41586-021-03819-2 Jul 15, 2021

20. Kuhn, Thomas S. The Structure of Scientific Revolutions. Chicago: University of Chicago Press, 1962.

21. Leonelli, S. (2015). What counts as scientific data? A relational framework. Philosophy of Science, 82(5), 810-821. January 1, 2022, Cambridge University Press, https://doi.org/10.1086/684083

22. Leonelli, S. (2019 a). "Data — from objects to assets," Nature, Nature, vol. 574(7778), pages 317-320, October.

23. Leonelli, S. (2019a). What distinguishes data from models? European Journal for the Philosophy of Science, 9(2), Article 22. https://doi.org/10.1007/s13194-018-0246-0

24. Leonelli, S. (2019b). "Data Governance is Key to Interpretation: Reconceptualizing Data in Data Science" Harvard Data Science Review, 1(1). https://doi.org/10.1162/99608f92.17405bb6

25. Leslie, David. 'Tackling COVID-19 Through Responsible AI Innovation: Five Steps in the Right Direction'. Harvard Data Science Review, no. Special Issue

26. Makmun, Abu Hassan. (2020). On the quality of qualitative research: a simple self-reminder. 10.13140/RG.2.2.35384.98565.

27. Makmun, Abu Hassan. (2020). Research paradigm (presentation). 10.13140/RG.2.2.10638.59202.

28. Mendling, J., Leopold, H., Meyerhenke, H., & Depaire, B. (2023). Methodology of Algorithm Engineering. ArXiv, abs/2310.18979.

29. Miller KL et al. Multimodal population brain imaging in the UK Biobank prospective epidemiological study. Nat.

29. Neurosci. 19, 1523–1536 (2016). [PubMed: 27643430]

30. Paunović, K. (2008). Data, Information, Knowledge. In: Kirch, W. (eds) Encyclopedia of Public Health. Springer, Dordrecht. https://doi.org/10.1007/978-1-4020-5614-7_685

31. Pearl, Judea, and Dana Mackenzie. 2019. The Book of Why. Harlow, England: Penguin Books.

32. Spiegelhalter, David. 'Should We Trust Algorithms?' Harvard Data Science Review 2

33. Stodden, Victoria. The Data Science Life Cycle: A Disciplined Approach to Advancing Data Science as a Science. Commun. ACM 63, no. 7 (2020): 58-66.

34. The world's largest health-research study is under way in Britain. It is aimed at saving Britons—and the NHS, The Economist, October 18, 2023.

35. Tukey, John W, "The Future of Data Analysis," The Annals of Mathematical Statistics 33, no. 1 (1962): 6,

36. Tukey, John, W, Exploratory Data Analysis, Addison Wesley, 1977

37. Van Essen DC et al. The WU-Minn Human Connectome Project: an overview. Neuroimage 80, 62– 79 (2013). [PubMed: 23684880]

38. Welsh, Matt, Large Language Models and The End of Programming, lecture at Harvard, October 24, 2023. https://www.fixie.ai

39. Varadi M, Anyango S, Deshpande M, Nair S, Natassia C, Yordanova G, Yuan D, Stroe O, Wood G, Laydon A, Žídek A, Green T, Tunyasuvunakool K, Petersen S, Jumper J, Clancy E, Green R, Vora A, Lutfi M, Figurnov M, Cowie A, Hobbs N, Kohli P, Kleywegt G, Birney E, Hassabis D, Velankar S. AlphaFold Protein Structure Database: massively expanding the structural coverage of protein-sequence space with high-accuracy models. Nucleic Acids Res. 2022 Jan 7;50(D1):D439-D444. doi: 10.1093/nar/gkab1061 PMID: 34791371; PMCID: PMC8728224.

40. Zhang, D., et. al., "The AI Index 2023 Annual Report," AI Index Steering Committee, Institute for Human-Centered AI, Stanford University, Stanford, CA, April 2023.